**The effect of the heating rate on the order to order phase transition**

Aycan Özkan, Bülent Kutlu

Gazi Üniversitesi, Fen -Edebiyat Fakültesi, Fizik Bölümü , 06500

Teknikokullar, Ankara, Turkey,

e-mail: aycan@gazi.edu.tr, bkutlu@gazi.edu.tr

### Abstract

The simple cubic spin-1 Ising (BEG) model exhibits the ferromagnetic ($F$) - ferromagnetic ($F$) phase transition at low temperature region for the interval $1.40 < d = D/J < 1.48$ at $k = K/J = -0.5$. The degree of the $F - F$ phase transition determines the special point on the ($kT/J$, $d$) phase diagram In this paper, the critical behavior of the ferromagnetic ($F$) - ferromagnetic ($F$) phase transition has been investigated for different heating rates using the cellular automaton (CA) heating algorithm. The variations with heating rate for the $F - F$ phase transition order are quite important at the creation of phase diagrams. Therefore, the universality class and the type of ferromagnetic ($F$) - ferromagnetic ($F$) phase transition have been researched using the finite - size scaling theory, the power law relations and the probability distributions. The results show that the $F - F$ phase transition can be the second order, the first order or the weak first order depending on the heating rate in the interval

$1.40 < d < 1.48$ for $k = -0.5$.

## 1 Introduction

In recent years, the spin-1 Ising model, which is known as the Blume-Emery-Griffiths (BEG) model, can be used to simulate many physical systems. The model firstly has been presented for describing phase separation and superfluid ordering in He mixtures [1]. The versions of the model have been applied to the physical systems such as $He^3$-$He^4$ mixtures [1], the solid-liquid-gas systems [2], the multicomponent fluids [3], the microemulsions [4], the semiconductor alloys [5 − 7] and the binary alloys [8].

The Hamiltonian of the BEG model is given by

$$H_I = -J \sum_{<ij>} S_i S_j - K \sum_{<ij>} S_i^2 S_j^2 + D \sum_i S_i^2 \qquad (1)$$

where $\langle ij \rangle$ denotes summation over all nearest-neighbor pairs of sites and $S_i = -1, 0, 1$. The parameters $J$, $K$, and $D$ are bilinear, biquadratic interaction terms and the single-ion anisropy constant. The BEG model has been extensively studied by different techniques; the molecular field approximation (MFA) [1 − 3, 5, 9], Bethe approximation [4], Kikuchi approximation [6, 7], the mean field approximation (MFA) [10], Monte Carlo method (MC) [7, 8, 11], the transfer matrix method [12], series expansion method [13], the coupling constant approximation [14], the position space renormalization group method (PSRG) [15], cluster variation method (CVM) [16], linear-chain approximation [17], the path probability method (PPM) [18], Monte Carlo Renormalization Group Theory (MCRG) [19] and Cellular Automaton (CA) [20 − 30]. These studies show that the phase diagrams of the BEG model include a wide variety features. Primarily, the ground state phase diagrams of the BEG model have the perfect zero (PZ), antiquadrupole (a), saturated ferromagnetic and ferrimagnetic (f) ordering regions as well as ferromagnetic (F), antiferromagnetic (AF) and paramagnetic (P) ordering regions unlike the spin-1/2 Ising model [10 − 29, 32 − 36]. On the other hand, the model exhibits successive, re-entrant and double re-entrant phase transitions in the different regions of the parameter space. As a result, it is found that the phase diagrams have different special points named as critical point, tricritical point, critical end point, tetracritical point, multicritical and bicritical point [10 − 29, 32 − 36]. The special points are determined according to positions and degrees of phase transition lines. The phase transition lines correspond to first and second order phase transitions. As it is known, first and second order phase transitions are distinguished by the behavior of the thermodynamic quantities near the transition point temperature. First order phase transitions exhibit a wide variety of behaviors. In some of first order transitions, thermodynamic quantities exhibit s-shape or the discontinuity behavior at the transition point. On the other hand, some first order phase transitions appear like a second order one at the transition point and transitions are named as weak first order [31,32]. In this transitions, critical exponents have different values from the universal values. The creation of the phase diagrams of the BEG model is quite difficult due to variety in nature of the first order transition. This situation explains the differences between the results of the previous studies for the $(kT/J, d)$ phase diagrams through the $k$ line [10, 19 − 21, 32 − 34]. For example, the special point on the $(kT/J, d)$ phase diagram through the $k = −0.5$ line has determined as the critical end point (CEP) by the MFA [10], MCRG [19] and RG [34], the tricritical point (TCP) by CA [20], CVM [33] and TPCA [32] for the simple cubic lattice $(sc)$ and the bicrital point by CA[21] for the face centered cubic lattice $(fcc)$. These difference arises from changes in the phase transition type for every model in the interval $1.40 < d < 1.46$ The aim of this paper is to determine the causes of the difference occurring in the phase diagram of the simple cubic lattice $(sc)$. Considering the phase diagrams generated for $k = -0.5$, there are the successive $(F - P - F - F - P)$ phase transitions for MFA, the successive $(F - F - P)$ phase transitions for RG and the double re-entrant $(F - P - F - P)$ phase transitions for TPCA around the end (E) point. The type of the special point in this phase diagrams varies with the degree of phase transitions. The previous studies have shown that heating or cooling rate is very important in the arise of metastable states which cause the first-order phase transition [22].Therefore, in this paper, the $(kT/J, d)$ phase diagram for $k = -0.5$ has been obtained for procedures corresponding to different heating rates and the change of critical behavior depending on the heating rate has been investigated at different points of the phase diagram.

The order of the $F - F$ transition at the successive $(F - F - P)$ phase transition is very important for determining the phase boundary in the interval $1.4 < d < 1.50$ for the $k = -0.5$. Because these points create the type of the special point of the $(kT/J, d)$ phase space for $k = -0.5$. The degree of a point on the phase transition line can be determined with the behavior of thermodynamic quantities and the values of critical exponents. The obtained values of critical exponents must be compatible with the universal values ($\alpha = 0.12$, $\beta = 0.31$, $\gamma = 1.25$ and $\nu = 0.64$) for second order phase transition in 3-dimensional BEG model. The critical temperature and the statical critical exponents can be estimated by analyzing the data within the framework of the finite - size scaling theory and the power law relations [22].

In this paper, the simple cubic BEG model for $k = K/J = -0.5$ is simulated using heating algorithm improved from Creutz Cellular Automaton. In the previous papers, the Creutz cellular automaton (CCA) algorithm and its improved versions have been used successfully to study the properties of the critical behaviors of the Ising model Hamiltonians [20 − 29, 37 − 45]. The CCA algorithm, which was first introduced by Creutz [45], is a microcanonical algorithm interpolating between the conventional Monte Carlo and the molecular dynamics techniques. The Creutz cellular automaton (CCA) is faster than the conventional Monte Carlo method (MC). The CCA does not need high quality random numbers and it is a new and an alternative simulation method for physical systems. Our previous studies showed that the heating and the cooling algorithms improved from the Creutz cellular automaton algorithm are effective to study the phase space and the critical behavior of the Blume Emery Griffiths model [20 − 29].

## 2 Model

Three variables are associated with each site of the lattice. The value of each site is determined from its value and those of its nearest- neighbors at the previous time step. The updating rule, which defines a cellular automaton, is as follows: Of the three variables on each site, the first one is the Ising spin $B_i$. Its value may be 0 or 1 or 2 . The Ising spin energy for the model is given by Eq. (1). In Eq. (1), $S_i = B_i - 1$. The second variable is for the momentum variable conjugate to the spin ( the demon ). The kinetic energy associated with the demon, $H_K$, is an integer, which equal to the change in the Ising spin energy for the any spin flip and its values lie in the interval $(0, m)$. The upper limit of the interval, $m$, is equal to $24J$. During the simulation, the total energy

$$H = H_I + H_K \qquad (2)$$

is conserved.

The third variable provides a checkerboard style updating, and so it allows the simulation of the Ising model on a cellular automaton. The black sites of the checkerboard are updated and then their color is changed into white; white sites are changed into black without being updated. The updating rules for the spin and the momentum variables are as follows: For a site to be updated its spin is changed one of the other two states with 1/2 probability and the change in the Ising spin energy ,$dH_I$, is calculated. If this energy change is transferable to or from the momentum variable associated with this site, such that the total energy $H$ is conserved, then this change is done and the momentum is appropriately changed. Otherwise the spin and the momentum are not changed.

For a given total energy the system temperature is obtained from the average value of kinetic energy, which is given by:

$$\langle E \rangle = \frac{\sum_{n=0}^{m} n e^{-nJ/kT}}{\sum_{n=0}^{m} e^{-nJ/kT}} \qquad (3)$$

where $E = H_K$. The expectation value in Eq. (3) is average over the lattice and the number of time steps. Because of the third variable, the algorithm requires two time steps to give every spin of the lattice a chance to change. Thus, in comparison to ordinary Monte Carlo simulations, two steps correspond to one full sweep over the system variables.

The simulation has been run using heating algorithm [20, 21, 23 − 25]. The heating algorithm is divided into two basic parts, initialization procedure and the taking of measurements . In the initialization procedure, firstly, all spins in the lattice sites take the ferromagnetic ordered structure ( ↑ ↑ ) and the kinetic energy per site which is equal to the maximum change in the Ising spin energy for the any spin flip is given to the certain percent of the lattice using the second variables. This configuration is run during the 20.000 cellular automaton time steps. In the next steps, last configuration in the ordered structure has been chosen as a starting configuration for the heating run. Rather than resetting the starting configuration at each energy, it is used the final configuration at a given energy as the starting point for the next. During the heating cycle, a certain amount of energy per site has been added to the spin system through the second variables $(H_K)$ after the 2.000.000 cellular automaton steps. In this study, the heating rate corresponds to the rate of increase in kinetic energy$(H_k)$ for the lattice.

### 3 Results and Discussions

The all simulations have been done using the heating algorithm improved from CCA for the simple cubic BEG model [20]. The calculations have been repeated for different initial configurations and heating rates. Firstly, the$(kT/J, d)$ phase diagrams are constructed for rate of increasing $H_K$ which equal to $0.08H_k$ and $0.04H_k$ per site for $k = -0.5$. Afterwards, the calculations are repeated using three different procedures for every rate of increasing $H_k$. In these procedures, the heat is given to a certain percentage of the lattice points as it is seen in Table 1.

**Table 1** Heating procedures at the simulations

| Heating Rate | | |
|---|---|---|
| Procedure | Adding energy to a site | Percentage of the lattice site |
| I | $0.08 \times H_K$ | 100% |
| II | $0.08 \times H_K$ | 50% |
| III | $0.08 \times H_K$ | 25% |
| IV | $0.04 \times H_K$ | 100% |
| V | $0.04 \times H_K$ | 50% |
| VI | $0.04 \times H_K$ | 25% |

The computed values of the thermodynamic quantities ( the order para-meters ($M$, $Q$), the susceptibility ($\chi$), the spin-spin interaction energy ($U$), the specific heat ($C$) and Binder cumulant ($g_L$) ) are averages over the lattice and over the number of time steps (2.000.000 ) with discard of the first 100.000 time steps during the cellular automaton develops. They have been computed on the simple cubic lattice with $L = 12$, 16, 20, 24 and 32 for periodic boundary conditions.

The thermodynamic quantities are calculated from

$$M = \frac{1}{N} \sum_i S_i, \qquad Q = \frac{1}{N} \sum_i S_i^2 \tag{4}$$

$$U = -\frac{J}{U_0} \sum_{<ij>} S_i S_j \tag{5}$$

$$\chi = N \frac{\langle M^2 \rangle - \langle M \rangle^2}{kT} \tag{6}$$

$$C_I/k = N \frac{\langle U_I^2 \rangle - \langle U_I \rangle^2}{(kT)^2} \tag{7}$$

$$g_L = 1 - \frac{\langle M^4 \rangle}{\langle M^2 \rangle^2} \tag{8}$$

where $U_0$ is the ground state energy at $kT/J = 0$.

For $k = -0.5$, ground state phase diagram has ferromagnetic ($F$) and paramagnetic ($P$) phases. Ferromagnetic ($F$) and the paramagnetic ($P$) phases have been determined with the values of $M$ and $Q$ order parameters as [10],

Ferromagnetic $(F)$ : $M \neq Q \neq 0$

Paramagnetic $\quad (P)$ : $M = 0, Q \neq 0$.

Mainly, the phase transition type can be determined from the temperature dependence of thermodynamic quantities. The another procedure to distinguish the phase transition type is to calculate the probability distributions for the order parameter $(P(M))$ and spin-spin interaction energy $(P(U))$. In this study, the probability distributions are calculated by

$$P_L(M) \quad = \quad \frac{N_M}{N_{CCAS}} \qquad (9)$$

$$P_L(U) \quad = \quad \frac{N_U}{N_{CCAS}} \qquad (10)$$

where $N_{CCAS}$ is the total number of the cellular automaton time steps, and $N_M$ and $N_U$ are the numbers of times that magnetization $M$ and spin-spin interaction energy $U$ appears, respectively. The histogram with 200 bins are used for plotting the probability distributions of the magnetization. The probability distributions of the order parameter $(P(M))$ near the phase transition temperature have two peaks at the second order phase transitions and three peaks at the first order phase transitions. At the same time, the probability distributions of the spin-spin interaction energy $(P(U))$ have a single peak at the second order phase transitions and two peaks at the first order phase transitions [13, 29, 46, 47].

On the other hand, the critical exponents which determined the critical behavior have been estimated using the finite size scaling theory. The scaling relations of the Binder cumulant($U_L$), the order parameter ($M$), the susceptibility ($\chi$) and the specific heat ($C$) are given by

$$U_L = G^\circ(\varepsilon L^{1/\nu}). \tag{11}$$

$$M = L^{-\beta/\nu} X^\circ(\varepsilon L^{1/\nu}) \tag{12}$$

$$kT\chi = L^{\gamma/\nu} Y^\circ(\varepsilon L^{1/\nu}) \tag{13}$$

$$C = L^{a/\nu} Z^\circ(\varepsilon L^{1/\nu}). \tag{14}$$

The infinite lattice critical behavior must be asymptotically reproduced for large $x = \varepsilon L^{1/\nu}$, that is,

$$X^\circ(x) \propto A x^\beta \tag{15}$$

$$Y^\circ(x) \propto B x^{-\gamma} \tag{16}$$

$$Z^{\circ}(x) \propto Cx^{-\alpha}. \tag{17}$$

where the $\alpha$, $\beta$ $\gamma$and $\nu$ critical exponents are equal to 0.12, 0.31, 1.25 and 0.64, respectively, for 3-d Ising model and $\varepsilon =(T - T_C(\infty))/T_C(\infty)$. Near the critical temperature ($\varepsilon \to 0$), the scaling data must be located on a single curve for the second order phase transitions based on the finite-size scaling theory. The infinite lattice critical temperature $T_C(\infty)$ can be estimated using the expression below.

$$T_C(L) = T_C(\infty) + aL^{-1/\nu}. \tag{18}$$

## 3.1 The variation of the ( kT/J, d ) phase diagram with heating rate

In Fig. 1, the $(kT/J, d = D/J)$ phase diagrams have been illustrated for the $0.08 \times H_k$ value. The figure includes results for the three different procedures ($I$, $II$ and $III$). The phase diagrams exhibit a special point that combines three different lines These lines constitute the $F - P$ ($d < 1.44$ for procedure $I$ and $II$, $d < 1.41$ for procedure $III$, and $d > 1.5$ for all procedures), the successive $F - F - P$ ($1.44 \leq d < 1.48$) and the double reentrant $F - P - F - P$ ($1.48 \leq d \leq 1.5$) phase transitions. On the phase diagrams, the high temperature $F - P$ phase transitions and the $P - F$ phase transitions are of second order. These phase transitions generate the upper critical line on the phase diagrams. The lower temperature part of the double reentrant phase transition and the low temperature $F - P$ phase transitions generate the first order line ($d > 1.5$). On the other hand, the $F - F$ part of the successive $F - F - P$ phase transitions appear as the second order. The fluctuations for $F - F$ phase transitions are difficult to detect because they are quite small next to the fluctuations of other second order phase transitions. The $F - F$ phase transitions create another phase transition line which determines the type of the critical point. Therefore, the critical exponents for the $F - F$ phase transitions must be calculated to determine the degree of the phase transition. For this purpose, the values of critical exponents have been obtained using procedure $I$, $II$ and $III$ using the scaling relations of thermodynamic quantities for the branch consisting of the $F - F$ phase transitions. In Fig. 2, the finite-size scaling plots of the Binder cumulant ($U_L$), the order parameter ($M$), the susceptibility ($\chi$) and the specific heat ($C/k$) are shown at $d = 1.44$ over the branch consisting of the $F - F$ phase transitions for the procedure$III$. The finite lattice critical temperatures are obtained from the susceptibility and the specific heat maxima. The infinite lattice critical temperature has been estimated using Eq. (18) as $T_{C1}^{\chi}(\infty) = 0.69 \pm 0.01$ for $F - F$ phase transition at $d = 1.44$. The critical exponent $\nu$ can be obtained using the finite-size scaling relation given in Eq. (11) for the Binder cumulant. The scaling data of Binder cumulant near the infinite lattice critical temperature for the $L = 12, 16, 20, 24$ and $32$ finite-size lattices lie on a single curve for the universal value of $\nu = 0.64$ (Fig. 2a). The scaling data of the order parameter with $\beta = 0.31$ and $\nu = 0.64$ is shown in Fig. 2b together with straight line describing the theoretically predicted behavior for large $x$. The scaling data of the order parameter for $T < T_{C1}^{\chi}(\infty)$ are in good agreement with the asymptotic behavior given in Eq. (15). The slope of this straight line is equal to the universal value ($\beta = 0.31$) of the order parameter critical exponent. In Fig. 2c, the scaling data of susceptibility for $T < T_{C1}(\infty)$ with $\gamma = 1.25$ and $\nu = 0.64$ are shown together with straight line describing the theoretically predicted behavior given in Eq. (16) for large $x$. The slope of straight line passing through the susceptibility data is equal to the universal value ($\gamma = 1.25$). The specific heat on the infinite lattice is well described by

$$\frac{C}{k} = A\varepsilon^{-\alpha} + b^{\pm} \tag{19}$$

where $b$ express the nonsingular part of the specific heat. The finite-size scaling plot of the singular portion of the specific heat ($C/k - b^{-}$) is shown in Fig. 2d.

The data for $T < T_{C1}(\infty)$ of the specific heat are well scaled with $\alpha = 0.12$ and $\nu = 0.64$ for $b^- = -20$ (Eq. (14)). The scaling data of the specific heat are in good agreement with $\alpha = 0.12$ sloping straight line. Additionally, the $\alpha$, $\beta$, $\gamma$ and $\nu$ critical exponents are estimated for different $d$ values over the branch consisting of the $F-F$ phase transitions using the procedure $I$, $II$ and $III$. The estimated values of the critical exponents are consistent with universal values for second order phase transition. The values of the critical exponents have been confirmed using power law relations. Therefore, the branch consisting of the $F-F$ phase transitions is a second order critical line. Finally, there occurs the bicritical point ($BCP$) at $d = 1.48$ where the two critical lines meet with the first order phase transition line for all procedures as CA result [22]. On the other hand, the susceptibility peaks of the $F-F$ are much lower than the peaks of the $F-P$ phase transition. As it is seen in Fig. 3, the susceptibility peak of the $F-F$ phase transition may not be noticed for $d < 1.47$. In this situation, the special point can be obtained as $TCP$ instead $BCP$ as results of TPCA [32], CVM [33] and CA [20].

The phase diagrams for the procedures $IV$, $V$ and $VI$ corresponding to the $0.04 \times H_k$ value are illustrated in Fig. 4. These phase diagrams exhibit a different behavior from each other. The phase diagram created using procedure $IV$ has $F - P$ ($d < 1.44$ and $d > 1.5$), the successive $F - F - P$ ($1.44 \leq d < 1.47$) and the double reentrant $F - P - F - P$ ($1.47 \leq d \leq 1.5$) phase transitions. The estimated phase diagram for the procedure $IV$ is similar with phase diagrams of the procedures $I$, $II$ and $III$. As it is seen in Fig. 5, the Binder cumulant ($U_L$), the order parameter ($M$), the susceptibility ($\chi$) and the specific heat ($C/k$) data are scaled well using the universal critical exponent values at $T < T_{C1}(\infty)$ in the interval $1.44 \leq d < 1.47$ for procedure $IV$. Therefore, the phase diagrams for procedure $IV$ has a $BCP$ at $d = 1.47$.

On the other hand, it is seen in Fig. 4 that there occur the $F - P$ ($d < 1.4$ and $d > 1.5$), the successive $F - F - P$ ($1.4 \leq d < 1.44$) and the double reentrant ($1.44 \leq d \leq 1.5$) phase transitions for procedure $V$ and $VI$. Interestingly, the $F - F - P$ phase transitions in the interval $1.44 \leq d \leq 1.47$ for procedure $IV$ convert to the double reentrant $F - P - F - P$ phase transitions for procedure $V$ and $VI$ depending on the heating rate. Furthermore, the $F - P - F$ part of the double reentrant phase transition is of the first order. The phase diagrams for procedure $V$ and $VI$ have a special point at $d = 1.44$. The type of the special point are determined by degree of the $F - F$ phase transition in the interval $1.4 \leq d < 1.44$. For the special point at $d = 1.44$, the temperature dependences of the order parameter ($M$) and the susceptibility ($\chi$) are given for different procedures in Fig. 6 . It is seen in Fig. 6 that the successive $F - F - P$ phase transitions are of the second order since the order parameter shows continuity and the susceptibility has two characteristic peaks at the critical temperatures $(T_{C1}^{\chi}(\infty)$ and $T_{C2}^{\chi}(\infty))$ for procedure $I$ and $IV$. On the other hand, the $F-P-F$ parts of the double-reentrant phase transition are of the first order for procedure $V$. The behavior of the order parameter appears discontinuous at $F - P - F$ parts and continuous at high temperature $F - P$ part of the double reentrant phase transition. The susceptibility $(\chi)$ has two sharp peaks instead of a broad peak corresponding to the $F - F$ phase transition (Fig. 6).

The conversion of the phase transition from $F-F$ to $F-P-F$ is possible to detect using probability distributions. For the procedure $IV$, the $P(M)$ probability distribution exhibits a single peak for $F - F$ phase transition at $d = 1.44$. However, the $P(M)$ has two peaks for the low temperature $F-P$ and $P - F$ parts of the double reentrant phase transition due to the emergence of metastable state for procedure $V$ (Fig. 7(a) and 7(b)). In this histogram, the localized peak at $M = 0$ indicates paramagnetic $(P)$ phase while the peaks at $\pm M$ show ferromagnetic $(F)$ phase. The coexistence of ferrimagnetic $(F)$ and paramagnetic $(P)$ phases proves the first order phase transition. Similarly, the probability distribution of the Ising energy $(P(U))$ shows the double peaks at the transition temperatures(Fig. 7(c) and 7(d)). This case shows that the low temperature $F - P$ and $P - F$ parts of the double reentrant phase transition are of the first order for procedure $V$ at $d = 1.44$. It is clear that this change due to the heating rate. For the determined of the special point at $d = 1.44$, the values of the critical exponents for the $F - F$ phase transitions must be calculated for procedure $V$ and $VI$.

For analysis of the order of the $F - F$ phase transition on the phase diagrams created for the procedure $V$ and $VI$, the data of the thermodynamic quantities are scaled using finite-size scaling for $d = 1.43$ in Fig. 8. The correlation length critical exponent is estimated as $\nu = 0.75$ using the Binder cumulant scaling relation (Fig. (8a)). The scaling data of the order parameter, the susceptibility and the specific heat lie on the straight lines with the $\beta = 0.11$, $\gamma = -0.5$ and $\alpha = -0.07$ slopes describing the theoretically predicted asymptotically behaviors for large $x$, respectively. These estimated critical exponent values are nonuniversal (Fig. 8). These results are confirmed by using power law relations below.

$$M(L) \propto \varepsilon^{\beta(L)} \tag{20}$$

$$\chi(L) \propto \varepsilon^{-\gamma(L)} \qquad (21)$$

$$C(L) \propto \varepsilon^{-\alpha(L)} \qquad (22)$$

where $\varepsilon = (T - T_C(L))/T_C(L)$. The finite lattice critical exponents $\beta(L)$, $\gamma(L)$, and $\alpha(L)$ of the order parameter $(M)$, susceptibility $(\chi)$ and the specific heat $(C)$ quantities are obtained from the slope of the log-log plot of the power laws relations for each finite lattices. As it is seen in Fig. 9, the critical exponent values are obtained as $\beta = 0.113 \pm 0.001$, $\gamma = -0.491 \pm 0.001$ and $\alpha = -0.070 \pm 0.011$ at $d = 1.43$ for procedure $V$. These values are same with results of the finite-size scaling analysis. Both analysis results indicated that the $(F - F)$ phase transition which appears continuous is not of the second order. The reason of this nonuniversal behavior is the metastable states which arise using different heating procedures. At the points over the branch consisting of the $F - F$ phase transitions, the values of the critical exponents are different from universal values ( $\alpha = 0.12$, $\beta = 0.31$, $\gamma = 1.25$ and $\nu = 0.64$) for second order phase transition. So, the $F - F$ phase transitions is of the weak first order for $d = 1.43$ . Our analyses show that, there is the weak first order phase transition line corresponding to the $F - F$ phase transitions in the interval $1.4 \leq d < 1.44$.

At $d = 1.44$, the second order line corresponding to high temperature $F - P$ and $P - F$ phase transitions terminates on the first order line generated by the weak first order $F - F$ and the first order $F - P$ phase transitions. Thus, there occurs the critical end point ($CEP$) on phase diagram for procedure $V$ and $VI$ as results of the MFA [10], MCRG [19] and RG [34].

### 4 Conclusion

The results indicate that the algorithm has a major impact on the simulation of BEG model. As it is known, the metastable states leads to emergence of the first order phase transition. The details of the algorithm and heating rate are very effective in the arising of metastable states. Therefore, the $(kT/J, d)$ phase diagrams varies according to the heating rate. As shown, a phase transition of the second order for $0.08 \times H_k$ consists of first order for $0.04 \times H_k$.at $d = 1.44$. The result of simulations performed using CCA, the phase diagram exhibited two different special points as the bicritical point ($BCP$) at $d = 1.47$ and the critical end point ($CEP$) at $d = 1.44$ depending on the procedures for $k = -0.5$. In these simulations, some of the procedures can not produce metastable states. Therefore, $BCP$ is obtained instead of $CEP$. Our results indicate that the special point on the $(kT/J, d)$ phase diagram for $k = -0.5$ can be changed depending on details of the algorithms and heating rate. This consequence explains the differences between the results of the previous studies for the $(kT/J, d)$ phase diagrams at $k = -0.5$ [10, 19 − 21, 32 − 34].

Automaton. Chineese Phys. Lett. 23, 2526-2529 (2006)

**Figure Captions**

**Fig. 1** The $(kT/J, d)$ phase diagrams at $k = -0.5$, for procedure $I$, $II$ and $III$. The dashed and solid lines present the first order and the second order phase transitions, respectively

**Fig. 2** For procedure $III$, the finite size scaling plots of **(a)** the Binder cumulant $(U_L)$, **(b)** the order parameter $(M)$, **(c)** the susceptibility $(\chi)$ and **(d)** the specific heat $(C/k)$

**Fig. 3** The susceptibility $(\chi)$ curve for $d = 1.4$, $1.44$, $1.45$ and $1.46$ values for procedure I

**Fig. 4** The $(kT/J, d)$ phase diagrams at $k = -0.5$, for procedure $IV$, $V$

and $VI$. The dashed and solid lines present the first order and the second order phase transitions, respectively

**Fig. 5** For procedure $IV$, the finite size scaling plots of **(a)** the Binder cumulant $(U_L)$, **(b)** the order parameter $(M)$, **(c)** the susceptibility $(\chi)$ and **(d)** the specific heat $(C/k)$ at $d = 1.44$

**Fig. 6** For procedure $IV$ and $V$, the temperature dependence of **(a)** The order parameters $(M, Q)$ and **(b)** the susceptibility $(\chi)$ at $d = 1.44$

**Fig. 7** For procedure $V$, probability distributions of **(a)** the order parameter $(P(M))$ for lower temperature $F - P$ phase transition, **(b)** the order parameter $(P(M))$ for $P - F$ phase transition, **(c)** the spin-spin interaction energy $(P(U))$ for lower temperature $F - P$ phase transition and **(d)** the spin-spin interaction energy $(P(U))$ for $P - F$ phase transition at $d = 1.44$

**Fig. 8** For procedure $V$, the finite size scaling plots of **(a)** the Binder cumulant $(U_L)$, **(b)** the order parameter $(M)$, **(c)** the susceptibility $(\chi)$ and **(d)** the specific heat $(C/k)$ at $d = 1.43$

**Fig. 9** For procedure $V$, the plots of finite size critical exponents; **(a)** $\beta(L)$ against $L^{-1/\nu}$, **(b)** $\gamma(L)$ against $L^{-1/\nu}$ and **(c)** $\alpha(L)$ against $L^{-1/\nu}$ using $\nu = 0.75$ at $d = 1.43$

.

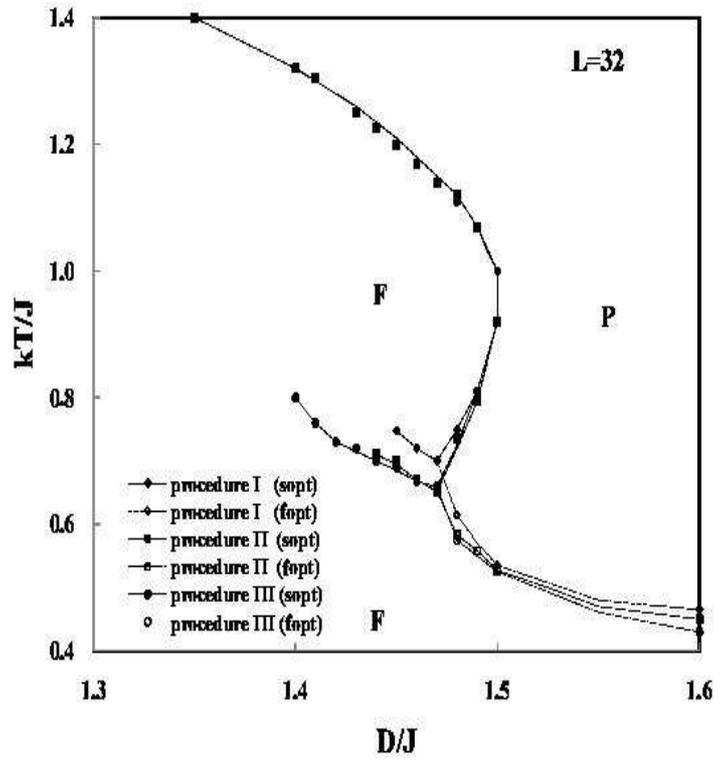

Figure 1: The $(kT/J, d)$ phase diagrams at $k = -0.5$, for procedure $I$, $II$ and $III$. The dashed and solid lines present the first order and the second order phase transitions, respectively

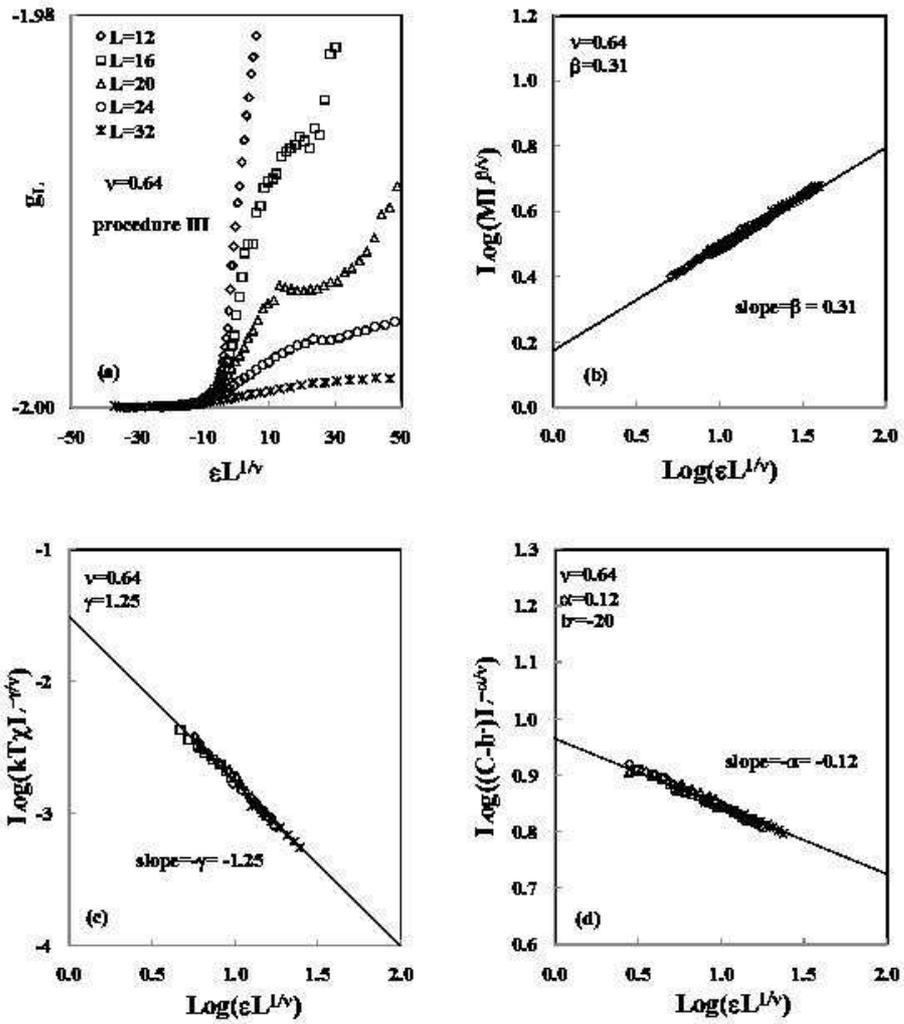

Figure 2: For procedure $III$, the finite size scaling plots of **(a)** the Binder cumulant ($U_L$), **(b)** the order parameter ($M$), **(c)** the susceptibility ($\chi$) and **(d)** the specific heat ($C/k$)

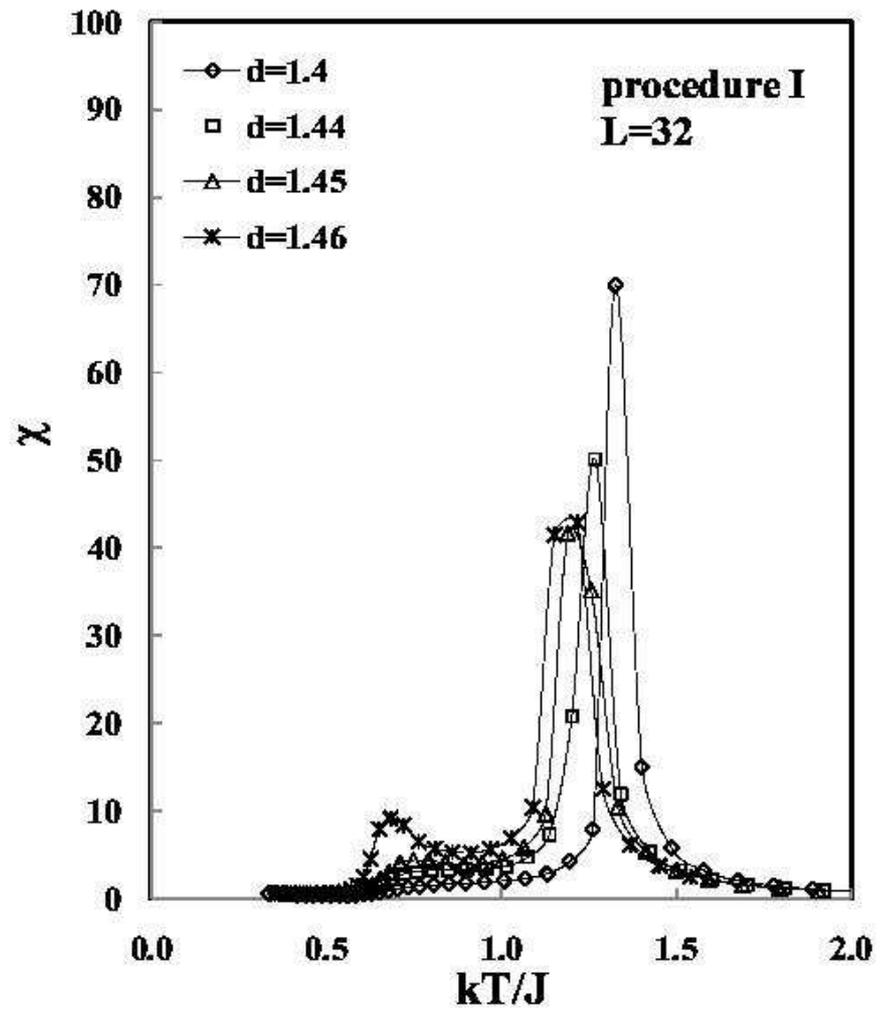

Figure 3: The susceptibility ($\chi$) curve for $d = 1.4$, 1.44, 1.45 and 1.46 values for procedure I

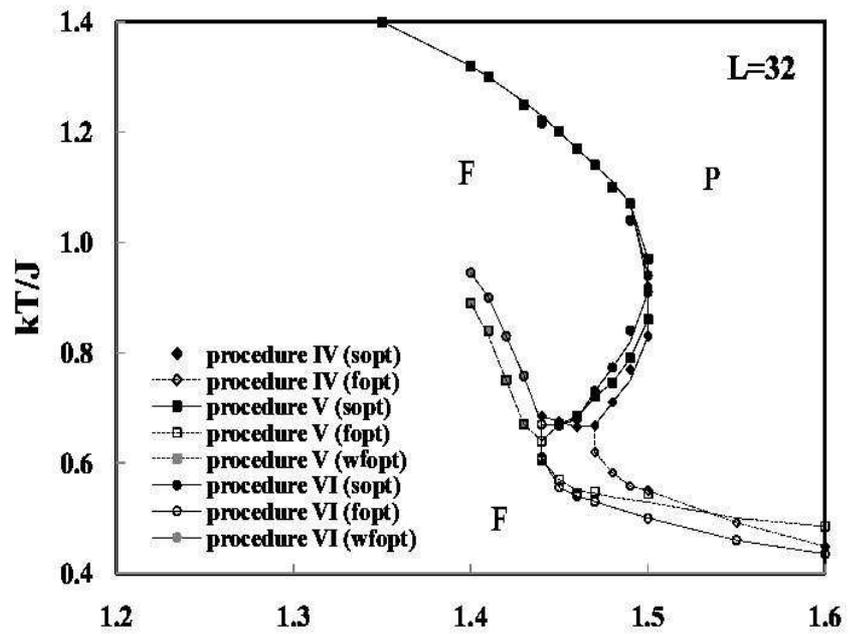

Figure 4: The $(kT/J, d)$ phase diagrams at $k = -0.5$, for procedure $IV$, $V$ and $VI$. The dashed and solid lines present the first order and the secon order phase transitions, respectively.

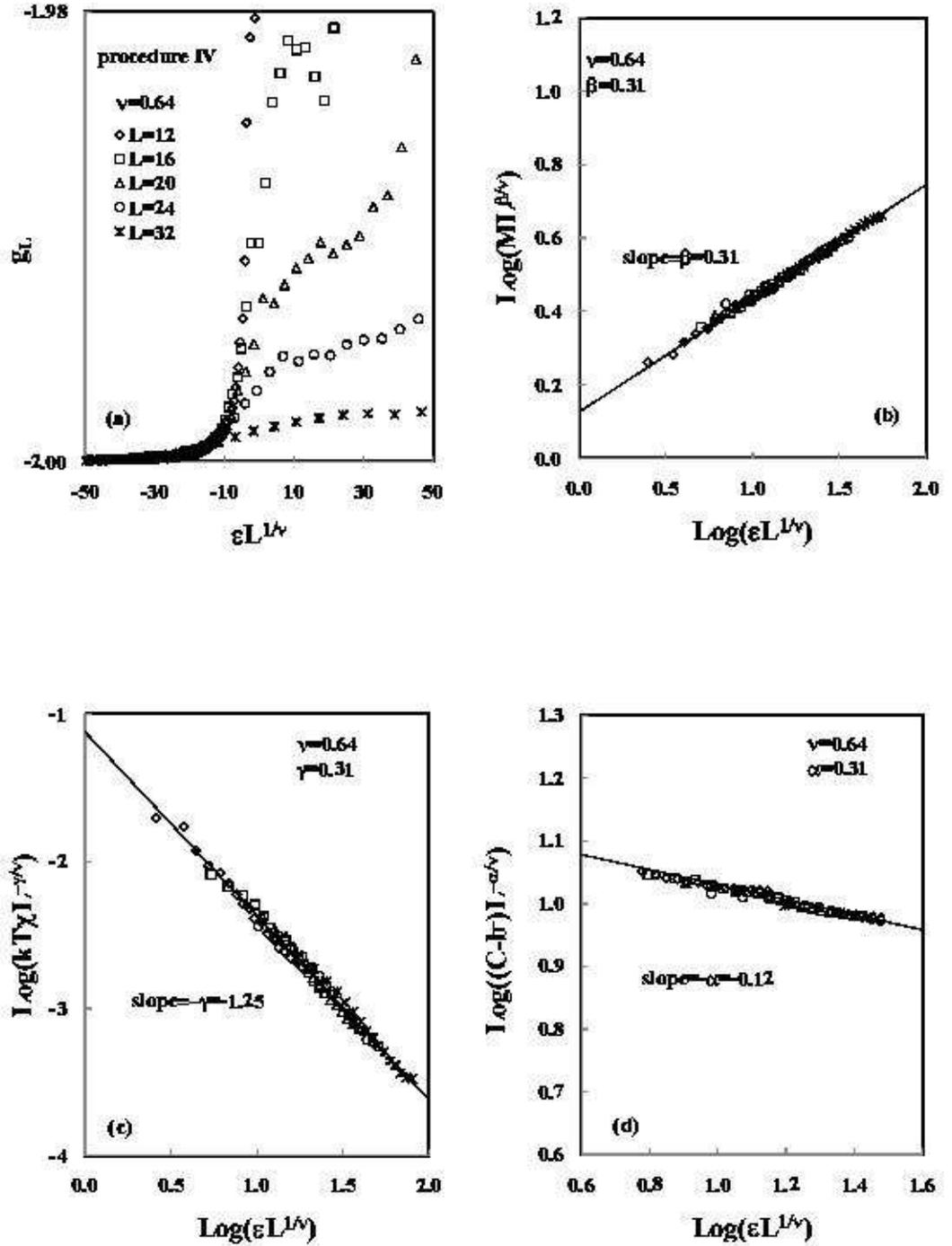

Figure 5: For procedure $IV$, the finite size scaling plots of **(a)** the Binder cumulant ($U_L$), **(b)** the order parameter ($M$), **(c)** the susceptibility ($\chi$) and **(d)** the specific heat ($C/k$) at $d = 1.44$

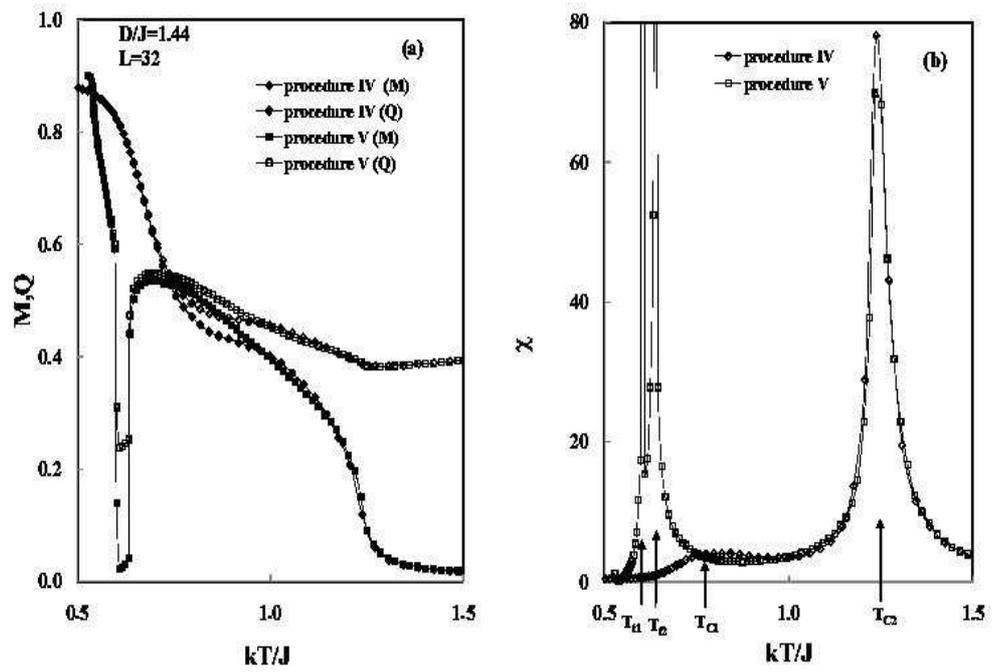

Figure 6: For procedure $IV$ and $V$, the temperature dependence of **(a)** The order parameters $(M, Q)$ and **(b)** the susceptibility $(\chi)$ at $d = 1.44$

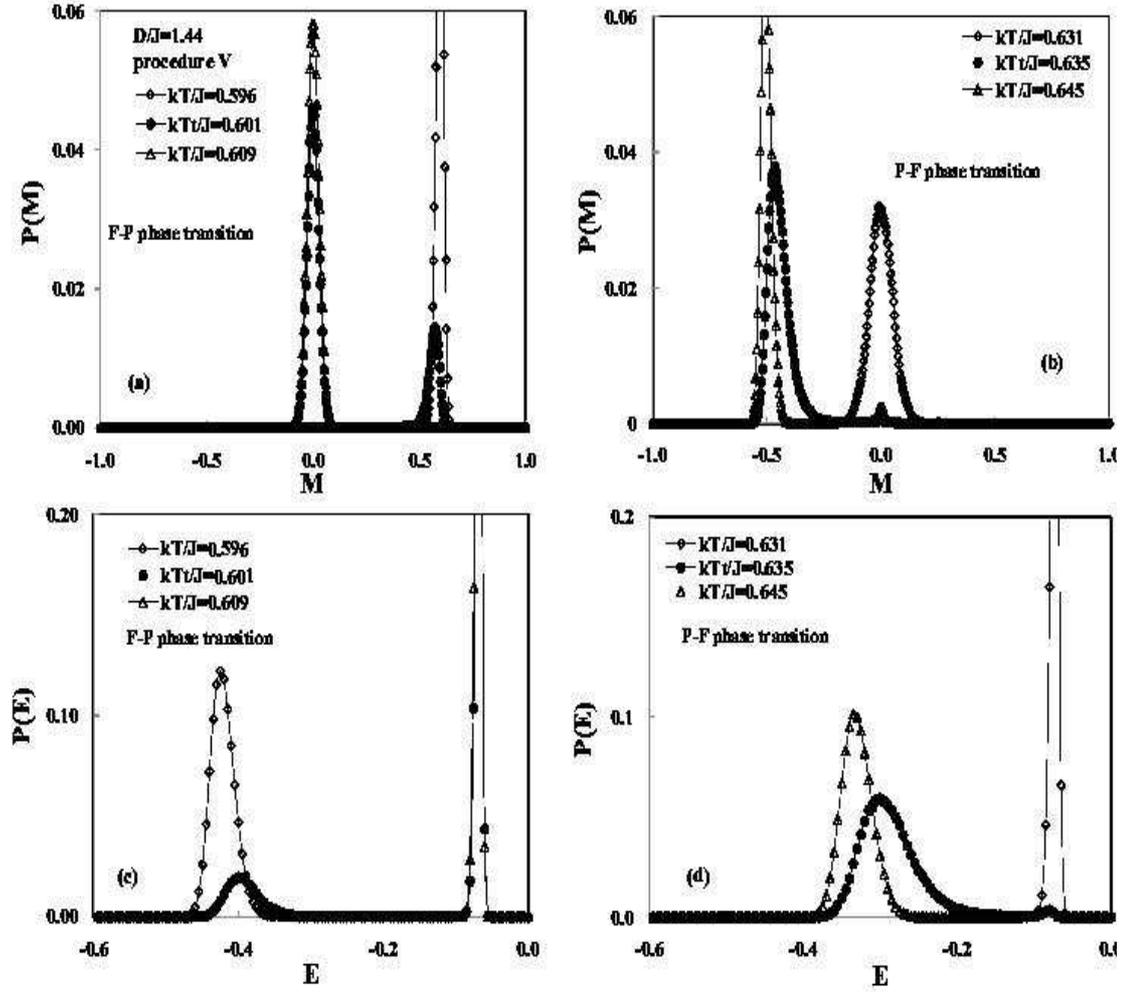

Figure 7: For procedure $V$, probability distributions of **(a)** the order parameter $(P(M))$ for lower temperature $F - P$ phase transition, **(b)** the order parameter $(P(M))$ for $P - F$ phase transition, **(c)** the spin-spin interaction energy $(P(U))$ for lower temperature $F - P$ phase transition and **(d)** the spin-spin interaction energy $(P(U))$ for $P - F$ phase transition at $d = 1.44$

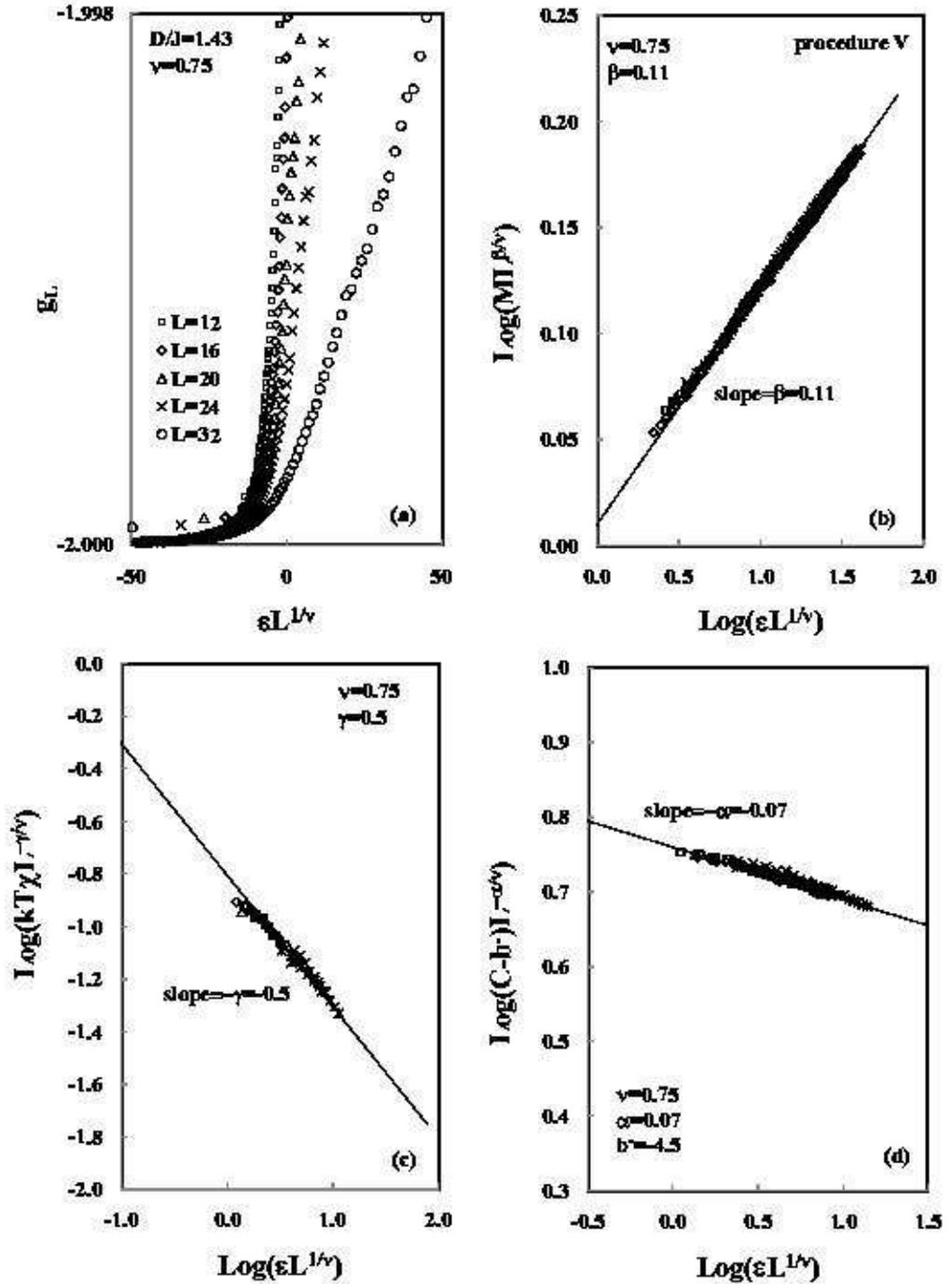

Figure 8: For procedure $V$, the finite size scaling plots of **(a)** the Binder cumulant ($U_L$), **(b)** the order parameter ($M$), **(c)** the susceptibility ($\chi$) and **(d)** the specific heat ($C/k$) at $d = 1.43$

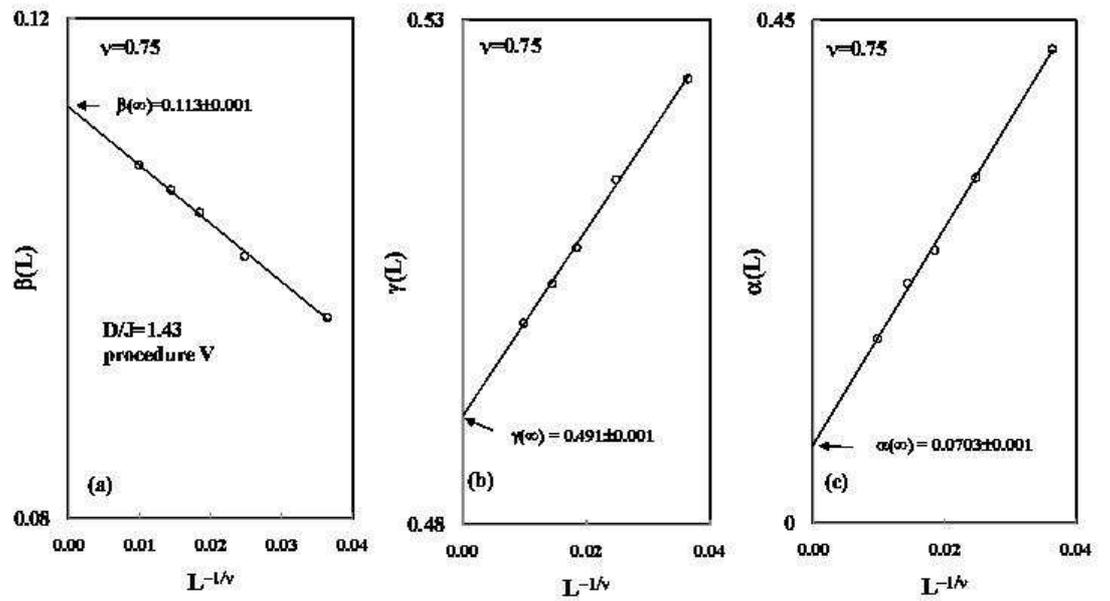

Figure 9: For procedure $V$, the plots of finite size critical exponents; **(a)** $\beta(L)$ against $L^{-1/\nu}$, **(b)** $\gamma(L)$ against $L^{-1/\nu}$ and **(c)** $\alpha(L)$ against $L^{-1/\nu}$ using $\nu = 0.75$ at $d = 1.43$